\documentclass[runningheads]{llncs}

\usepackage[T1]{fontenc}
\usepackage[utf8]{inputenc}
\usepackage{amsmath,amssymb}
\usepackage{mathtools}
\usepackage{graphicx}
\usepackage{booktabs}
\usepackage{algorithm}
\usepackage{algorithmic}
\usepackage{hyperref}
\usepackage{multirow}
\usepackage{color}
\usepackage{enumitem}
\usepackage{capt-of}
\usepackage{wrapfig}
\newcommand{\RR}{\mathbb{R}}
\newcommand{\E}{\mathbb{E}}
\newcommand{\norm}[1]{\left\|#1\right\|}
\newcommand{\abs}[1]{\left|#1\right|}
\newcommand{\BR}{\mathrm{BR}}

\begin{document}

\title{BRAID: Learning Equilibrium Maps in Interdependent Security Games
       via Weight-Tied Iterative Graph Neural Networks}
\titlerunning{BRAID: Learning Equilibrium Maps in IDS Games}
\author{
Elnaz Nowrouzi\inst{1} \and
Zhiqun Zuo\inst{1} \and
Xueru Zhang\inst{1} \and
Mohammad Mahdi Khalili\inst{1}
}

\authorrunning{E. Nowrouzi et al.}

\institute{
Computer Science and Engineering,
The Ohio State University\\
\email{
\{nowrouzi.1, zuo.167, zhang.12807, khalili.17\}@osu.edu
}
}

\maketitle

% ------------------------------------------------------------------
\begin{abstract}
Computing Nash equilibria in interdependent security (IDS) games on
networks is computationally expensive: best-response dynamics may need hundreds of
iterations per instance, and downstream tasks such as auditing,
stress-testing, and incentive design often require repeatedly re-solving the game under parameter perturbations.
We propose BRAID, a Best-Response Amortized Iterative Dynamics model that uses a
\emph{weight-tied iterative graph neural network} to learn a direct map from game parameters to Nash equilibrium effort profiles, replacing iterative best-response computation with a single forward pass that is up to $43\times$ faster per instance. BRAID is derived from the best-response fixed-point structure of IDS games: its SUM aggregation reflects additive neighbor coupling, and a weight-tied gated recurrent unit (GRU) mirrors a damped best-response update. The same architecture applies across IDS specifications that vary investment-cost curvature and neighborhood aggregation, including log-linear, quadratic-cost, and log constant-elasticity-of-substitution (CES) utilities. %The same architecture applies across a parameterized family of utilities spanning log-linear, quadratic-cost, and log constant-elasticity-of-substitution (CES) utilities. 
Beyond equilibrium prediction, BRAID also recovers how equilibrium efforts change under perturbations to game parameters, including costs and network edge weights. We make this sensitivity recovery an explicit evaluation target and introduce two training strategies, \textit{interior-equilibrium training} and \textit{input-noise regularization}, that improve the local behavior of the learned equilibrium map without using sensitivity labels. Experiments show that BRAID effectively predicts Nash equilibria and recovers equilibrium sensitivities across utility specifications and network sizes.
\end{abstract}

% ==================================================================
\section{Introduction}
% ==================================================================

Interdependent security (IDS) games model strategic environments in
which each agent's security outcome depends not only on its own effort, but also on the efforts of connected agents \cite{ballester2006s,jackson2008social,galeotti2010network,laszka2014survey,khalili2019public,khalili2020resource}.
Such interdependence is central to many security domains:
enterprise defenses are shaped by shared tenancy and supply-chain
dependencies \cite{khalili2018designing,khalili2019embracing,khalili2017designing,khalili2017embracing}; epidemic-prevention investments create spillovers across
contacts; and critical infrastructure often relies on pooled or mutually
reinforcing protection \cite{khalili2019incentivizing,hota2018interdependent}.
In such systems, each agent chooses a security effort to maximize its
own payoff while potentially benefiting from the efforts of its
neighbors. The Nash equilibrium (NE) is therefore the natural solution
concept: it describes a stable effort profile from which no agent can
unilaterally improve. It is also the key object for system-level
security analysis, since equilibria allow analysts to measure the
inefficiency of decentralized investment, identify structurally
under-protected agents, and design incentives  that steer the network toward safer
outcomes \cite{khalili2019incentivizing,hasheminasab2019design}.

Despite their usefulness, Nash equilibria are difficult to deploy as an
operational analysis tool. First, equilibrium computation can be costly. In general games, computing an NE is PPAD-complete, even for two-player games \cite{chen2006settling,daskalakis2009complexity}. In IDS network games, a standard approach is best-response dynamics (BRD), in which agents iteratively update their efforts by playing best responses to the current efforts of their neighbors. Although BRD exploits the game's structure, it may require many iterations to converge, and its per-instance cost grows with the number of agents \cite{ballester2006s,bramoulle2014strategic}.
Second, many security questions require not only an equilibrium, but its
response to parameter perturbations: how aggregate effort changes when an
agent's protection cost changes, how a stronger dependency reshapes the investment profile, or which parameter change most improves system-level safety.
Answering such questions with BRD requires re-solving the game for each perturbation, which becomes prohibitive when analysts need to audit, stress-test, or optimize over many game instances in near real time.

Learning-based equilibrium solvers offer a natural way to amortize equilibrium computation: after training on many solved game instances, they can predict equilibria for new instances with much lower inference cost. Existing approaches, however,  are not well suited to IDS network games. Neural Equilibrium Solvers \cite{marris2022turbocharging} rely on payoff-matrix representations, which scale poorly as the number of agents and actions grows. Nash Fixed-Point Networks
\cite{mckenzie2022operator} handle continuous games through learned fixed-point operators, but they treat each game instance in a generic form rather than explicitly exploiting the network structure that governs strategic interactions in IDS games. In IDS games, this relational structure is central: an agent's equilibrium effort is shaped not only by its own cost, but also by its position in the network and the efforts of its neighbors \cite{ballester2006s,bramoulle2014strategic,galeotti2010network}. While graph neural networks (GNNs) have been used for inverse problems such as recovering payoffs or network structure from observed equilibria
\cite{trivedi2020mine,rossi2022networkgames}, their use as direct, single-pass forward solvers for continuous-effort IDS games has received little attention.

A second gap concerns evaluation. Existing work typically reports only pointwise equilibrium error, but rarely evaluates whether the learned solver captures how equilibria respond to changes in game parameters. This matters because a solver can attain low prediction error at sampled instances while producing inaccurate local responses to perturbations, making it unreliable for what-if analysis, key-player identification, and mechanism design. We therefore argue that learned IDS solvers should be evaluated not only by equilibrium accuracy, but also by their ability to recover reliable parameter sensitivities.

We introduce BRAID, a Best-Response Amortized Iterative Dynamics model that uses a
weight-tied iterative graph neural network (GNN) to learn the equilibrium map
of IDS network games. At inference time, it serves as a single-forward-pass, amortized NE
solver, avoiding the need to run BRD separately for each new game
instance.
Its architecture is derived directly from the best-response fixed point: an agent's best response depends on a weighted aggregate of its neighbors' efforts, which corresponds naturally to SUM-aggregated message passing on the interaction graph;
 a  shared gated recurrent unit (GRU)
update cell plays the role of a damped best-response update; and tying
the parameters across rounds reflects the time-homogeneous structure of
BRD. One forward pass of BRAID therefore unrolls $K$ learned,
best-response-like updates using the same message and update operators
at each round. This design gives the model an inductive bias aligned
with the underlying game dynamics, rather than imposing a generic graph
architecture on the problem.

%This derivation does not depend on a particular closed-form utility. We instantiate and evaluate the \emph{same} architecture and training recipe on three IDS utility families with distinct best-response structure:
This derivation does not depend on a particular closed-form utility. We
instantiate and evaluate the \textit{same} architecture and training recipe on three
IDS specifications that vary two modeling dimensions: investment-cost
curvature and neighborhood aggregation. The log-linear utility provides an
additive baseline; the quadratic-cost model preserves the same aggregate
while introducing increasing marginal investment costs; and the log-CES
utility generalizes the additive aggregate by varying the substitutability
of neighboring efforts, with the log-linear case recovered as
$\rho\to1$. Our claim is therefore not breadth across unrelated utility
classes, but that one architecture, without utility-specific redesign,
spans these controlled variations in best-response structure. 

%(i) a log-linear utility, an instance of the linear-best-reply network games studied in~\cite{bramoulle2014strategic} and the local public-good form of~\cite{galeotti2010network}; (ii) the weighted-effort, quadratic-cost model of~\cite{khalili2019incentivizing}; and (iii) a log constant-elasticity-of-substitution (CES) utility~\cite{arrow1961capital}, that generalizes the additive aggregate of (i) along the standard CES substitutability axis. These are not unrelated forms: the log-linear utility is the $\rho\to1$ limit of log-CES, so the three trace a continuously parameterized family. Our claim is accordingly not the number of utilities, but that one architecture, with no utility-specific redesign, spans best-response maps from closed-form additive responses to a non-additive aggregate requiring a scalar root-find showing that BRAID captures the shared fixed-point structure of IDS games rather than an artifact of one utility form.

Our central claim goes beyond equilibrium prediction: BRAID can also
recover how equilibrium efforts change under perturbations to game
parameters, including costs and edge weights. We make sensitivity
recovery an explicit evaluation target and use two training strategies
that improve the local behavior of the learned equilibrium map without
requiring sensitivity labels. First, \emph{interior-equilibrium training} broadens the parameter
distribution and assigns a per-game coupling target so that equilibria
lie strictly inside the action space, where the best-response map is
smooth. This avoids boundary solutions, where the equilibrium response can be non-differentiable and sensitivities may be ill-defined. Second, \emph{input-noise regularization} perturbs node and edge
features during training, acting as a data-space smoothness prior that
penalizes large local derivatives in the output.
We validate the recovered sensitivities against ground truth obtained
from a symmetric finite-difference protocol in which each perturbed
equilibrium is recomputed by warm-started, re-verified BRD, avoiding
contamination from discontinuous jumps between equilibria. Our contributions are:
\begin{itemize}[leftmargin=*]
  \item \textbf{A best-response-derived architecture.}
    We introduce BRAID, a weight-tied graph neural network that is derived from the structure of damped best-response dynamics. BRAID uses SUM aggregation to capture weighted neighbor effects and a shared GRU update cell to emulate repeated best-response updates. This design yields a single-forward-pass, amortized NE solver whose inference is up to $43\times$ faster than iterative BRD per instance (Table~\ref{tab:runtime}).

  \item \textbf{Robustness across utility specifications.}
    We show that the same architecture and training recipe apply across a parameterized family of IDS utilities: log-linear, quadratic-cost, and log-CES, with the log-linear utility recovered as the $\rho\to1$ limit of log-CES. BRAID achieves relative equilibrium error below $0.5\%$ for the log-linear and quadratic-cost models and below $2\%$ for the more challenging log-CES model, with
    $R^2>0.999$ across all settings.

  \item \textbf{Sensitivity-aware learning and evaluation.}
We show that interior equilibrium training combined with input-noise
    regularization encourages a locally smooth learned equilibrium map
    whose derivatives recover equilibrium responses to parameter
    perturbations. Across the main settings, BRAID achieves $2$--$7\%$ relative error for cost sensitivities and $11$--$17\%$ relative error for edge-weight sensitivities on most settings, rising to $19$--$31\%$ for the most curved log-CES utility ($\rho{=}0.7$), all without ground-truth sensitivity supervision.
\end{itemize}

The rest of the paper is organized as follows.
Sec.~\ref{sec:related} reviews related work;
Sec.~\ref{sec:problem} formalizes the IDS game and research problem; Sec.~\ref{sec:model} derives BRAID from the
best-response fixed point and presents the architecture;
Sec.~\ref{sec:training} and Sec.~\ref{sec:sensitivity} describe the training procedure and sensitivity-recovery protocol, respectively.
Sec.~\ref{sec:experiments} and Sec.~\ref{sec:conclusion} present the experimental results, discussion, and conclusion.

% ==================================================================
\section{Related Work}
\label{sec:related}
% ==================================================================

\textbf{Computing Nash Equilibria.} Classical equilibrium computation relies on exact algorithms such as the
Lemke--Howson method \cite{lemke1964equilibrium} for bimatrix games and
support enumeration, both with exponential worst-case complexity.
This hardness is intrinsic: computing a NE is PPAD-complete
even for two-player games \cite{chen2006settling}, and for
the general problem \cite{daskalakis2009complexity}.
In the IDS games studied here (surveyed
in~\cite{laszka2014survey}), the standard tool is best-response
dynamics, which converges to a unique interior equilibrium under diagonal
strict concavity \cite{ballester2006s}. However, the number of iterations can grow with
network size and coupling strength. These  motivate approximate and
learning-based solvers that amortize equilibrium computation.

 \noindent \textbf{Learned Equilibrium Solvers.} Recent work has begun to cast equilibrium computation as a learning
problem. Neural Equilibrium Solvers (NES)
\cite{marris2022turbocharging} take the payoff tensor of a normal-form
game and output a Nash, correlated, or coarse-correlated equilibrium
(NE, CE, or CCE) in a single pass, using an architecture equivariant to
players and strategies. Its representation, however, is normal-form with
finite action sets, and does not encode an interaction network.
Nash Fixed-Point Networks (N-FPNs) \cite{mckenzie2022operator} address
\emph{contextual} games with unknown cost functions: they learn an
implicit operator whose fixed point is the variational-inequality
solution, and compute predictions by iterating that operator to
convergence at inference time, with the network entering through
constraints or context rather than a graph-aware predictor.
RENES \cite{wang2024renes} instead uses learning to improve classical
solvers: a reinforcement-learning policy modifies a normal-form game
before an existing inexact solver is applied. Although RENES uses a GNN,
its graph is an $\alpha$-rank response graph over joint action profiles
rather than the interaction topology among agents, and inference still
requires an inner solver run. A complementary, graph-based line studies the \emph{inverse} problem:
MINE \cite{trivedi2020mine} recovers agents' strategies and latent
rewards from observed networks via multi-agent inverse reinforcement
learning, and Rossi et al.~\cite{rossi2022networkgames} learn a map from
observed equilibrium actions to the underlying network structure. Both
infer games from behavior rather than predicting equilibria from known
games.
Closest in architectural spirit, Hu et al.~\cite{hu2025graph} develop
graph-structured architectures for equilibria of stochastic differential
games on graphs, sparsifying a feedforward network along the graph. We
share the premise that graph structure should be built into the
hypothesis class, but study static IDS games with continuous effort and
derive a weight-tied message-passing operator directly from the
best-response fixed point, so our unrolled depth is tied to best-response
dynamics itself.
Finally, Yu et al.~\cite{yu2024explore} approximate a \emph{weaker}
solution concept (CCE) online via Exp3-IX in stochastic cyber-defense
environments, learning through repeated interaction with a single
environment rather than amortizing an equilibrium map across a family of
games.

 \noindent \textbf{Network Games and GNNs for Optimization.} Game-theoretic foundations establish that network topology, not private
cost alone, drives equilibrium behavior: \cite{ballester2006s} identify
the ``key player'' whose structural position governs aggregate effort in
linear-quadratic network games, \cite{bramoulle2014strategic} show that
equilibria of linear-best-reply games are determined by the network's
lowest eigenvalue, and \cite{galeotti2010network} characterize how
position, the nature of strategic interaction, and information shape
behavior.
A complementary line tackles scale analytically: graphon games
\cite{parise2023graphon} approximate equilibria of large network games by
a limiting continuum model.
Within IDS games specifically, prior work has studied game design over
constrained influence networks \cite{hasheminasab2019design},
Stackelberg leader-follower variants \cite{huang2023stackelberg}, and
evolutionary dynamics on structured networks \cite{wang2021evolutionary}.
These motivate our own emphasis on a single architecture and training
recipe applied across network structures and sizes.
More broadly, GNNs are effective wherever the solution to a combinatorial
or physical problem depends on relational structure
\cite{battaglia2018relational}, building on standard message-passing
architectures \cite{kipf2017gcn,hamilton2017graphsage}; a related thread
treats a network's output as the fixed point of a tied operator, as in
deep equilibrium models \cite{bai2019deq}, of which our weight-tied
rounds are a finite unrolling (Sec.~\ref{sec:model}). We build on the
message-passing framework of \cite{scarselli2009graph} with GRU gating
\cite{li2015gated}.

 \noindent \textbf{Comparison with Prior Learned Solvers.} Table~\ref{tab:related} compares the closest prior learned equilibrium
solvers along five axes:
the \emph{game type} targeted; whether the method is \emph{graph-aware},
i.e., the interaction topology enters as an explicit input and inductive
bias rather than a flat game representation; the \emph{action space}
(IDS effort is continuous, so discrete-action methods require
discretization); the \emph{solution} concept produced; and whether a
prediction is \emph{single-pass}, i.e., one forward evaluation without an
inference-time iterative solve, which is what makes a solver genuinely
amortized. It highlights the gap addressed by this work: no prior method is
simultaneously a single-pass, graph-aware predictor of continuous-action
Nash equilibria in network games. NES and N-FPN treat a game as a
structureless object even though network position is the primary
determinant of IDS equilibrium effort; NES and the Exp3-IX learner are
confined to discrete actions; and only NES is single-pass, while MINE
solves the inverse problem. BRAID is designed to fill this gap;
Sec.~\ref{sec:model} discusses why direct adaptations of existing
methods do not provide the same combination of properties.

\begin{table}[!t]
\centering
\caption{Comparison of learned equilibrium prediction methods along the
  five axes defined in the text. ``Graph-aware'' marks an explicit
  dependence on interaction topology; ``Single-pass'' marks a prediction
  produced without an inference-time iterative solve. Entries for prior
  methods are drawn from their published descriptions.}
\label{tab:related}
\small
\setlength{\tabcolsep}{4.5pt}
\begin{tabular}{lccccc}
\toprule
\textbf{Method} & \textbf{Game type} & \textbf{Graph-} & \textbf{Action}
  & \textbf{Solution} & \textbf{Single-} \\
 & & \textbf{aware} & \textbf{space} & & \textbf{pass} \\
\midrule
NES~\cite{marris2022turbocharging}   & Normal-form        & \texttimes & Discrete & NE/CE/CCE & \checkmark \\
N-FPN~\cite{mckenzie2022operator}    & Contextual/convex  & \texttimes & Cont.\   & NE        & \texttimes \\
Exp3-IX$_{\text{rl}}$~\cite{yu2024explore} & Stochastic   & \texttimes & Discrete & CCE       & \texttimes \\
MINE~\cite{trivedi2020mine}          & Network (inverse)  & \checkmark & ---      & ---       & ---        \\
\textbf{BRAID}                       & \textbf{Network (forward)} & \checkmark & \textbf{Cont.} & \textbf{NE} & \checkmark \\
\bottomrule
\end{tabular}
\end{table}

% ==================================================================
\section{Problem Formulation}
\label{sec:problem}
% ==================================================================

\smallskip \noindent \textbf{Network Security Game.} Consider a network security game with $n$ agents, indexed by $i \in [n] \coloneqq
\{1,\ldots,n\}$, where each agent chooses an \textit{effort} level $e_i \in \RR_{\geq 0}$.
A directed dependency matrix $X=[X_{ij}]\in\RR_{\geq0}^{n\times n}$
encodes security spillovers: $X_{ii}>0$ denotes the self-benefit of agent $i$'s own effort,
normalized to $X_{ii}=1$ throughout, and $X_{ij}\geq0$ denotes the influence of
agent $j$'s effort on agent $i$.
Each agent $i$ has a \textit{cost} parameter $c_i>0$, which controls how costly
security effort is for that agent. Denote $c=(c_1,\ldots,c_n)$ as the vector of cost parameters.

Given a game instance $(X,c)$, each agent chooses $e_i$ to maximize its
\textit{utility} $U_i(e_i,e_{-i};X,c_i)$, which represents its security
benefit net of effort cost. The induced \textit{best-response map} is $\BR_i(e_{-i}) \in \arg\max_{e_i\in\RR_{\geq0}}
  U_i(e_i,e_{-i};X,c_i).$ A profile $e^*=(e_1^*,\ldots,e_n^*)$ is a Nash equilibrium (NE) if
$e_i^*=\BR_i(e_{-i}^*)$ for all $i\in[n]$. We focus on the parameter regime described in Sec.~\ref{sec:dataset}, under which the equilibrium is interior ($e_i^*>0$ for all $i\in[n]$) and, as we verify empirically in Sec.~\ref{sec:experiments}, unique across the sampled instances.

\smallskip \noindent \textbf{Three Utility Instances.}
\label{sec:utilities}
%We consider three families of IDS utilities, each combining a concave security benefit from pooled neighborhood effort with an individual investment cost. In each case, agent $i$'s best response depends on the efforts of other agents through a \textit{weighted neighbor aggregate}. Such concave benefits from weighted aggregate effort are standard in network-security and local-public-good games~\cite{ballester2006s,bramoulle2014strategic,galeotti2010network}: each agent's payoff increases in a weighted total of its own and its neighbors' effort net of cost, and the positive externality it induces is precisely what drives the under-investment that IDS analysis targets~\cite{khalili2019incentivizing}.
We consider three stylized IDS utility specifications chosen to vary two
modeling features that are central to security-investment games: the
curvature of investment costs and the way neighboring protection efforts
are aggregated. All three retain the standard IDS structure in which an
agent benefits from both its own security investment and positive
spillovers from neighboring agents, while bearing the cost of its own
effort~\cite{ballester2006s,bramoulle2014strategic,galeotti2010network,khalili2019incentivizing}.

The three specifications are not intended as unrelated utility
classes. Rather, (U1) serves as an additive baseline with linear investment
cost, (U2) preserves the same additive security aggregate but introduces
quadratic cost, and (U3) generalizes the aggregate in (U1) through a CES
form that varies the substitutability of neighboring efforts. This design
lets us test whether our solver remains effective when either the 
cost structure or the aggregation rule is changed.

\noindent \textit{(U1) Log-linear utility \cite{bramoulle2014strategic,galeotti2010network}.}
\begin{equation}
  U_i = \ln\!\Bigl(\textstyle\sum_{j} X_{ij}\,e_j\Bigr) - c_i e_i,
  \qquad
  \BR_i(e_{-i}) = \Bigl(\tfrac{1}{c_i}
    - \textstyle\sum_{j\neq i} X_{ij} e_j\Bigr)^{\!+},
  \label{eq:util_log}
\end{equation}
where $(z)^+=\max(0,z)$ and we  use the normalization $X_{ii}=1$.

 \noindent \textit{(U2) Quadratic-cost utility \cite{khalili2019incentivizing}.}
\begin{equation}
  U_i = \ln\!\Bigl(\textstyle\sum_{j} X_{ij}\,e_j\Bigr)
        - \tfrac{c_i}{2}\,e_i^2,
  \qquad
  \BR_i(e_{-i}) = \frac{-T_i + \sqrt{\,T_i^2 + 4X_{ii}^2/c_i\,}}{2X_{ii}},
  \label{eq:util_quad}
\end{equation}
with $T_i=\sum_{j\neq i} X_{ij} e_j$; the best response is the unique
non-negative root of the first-order condition
$c_i X_{ii} e_i^2 + c_i T_i e_i - X_{ii}=0$.

 \noindent \textit{(U3) Log-CES utility \cite{arrow1961capital}.}
\begin{equation}
  U_i = \tfrac{1}{\rho}\,
        \ln\!\Bigl(\textstyle\sum_{j} X_{ij}\,e_j^{\rho}\Bigr) - c_i e_i,
  \qquad \rho\in(0,1),
  \label{eq:util_ces}
\end{equation}
has interior best response solving
$e_i^{\rho-1}/(e_i^{\rho}+A_i)=c_i$ with
$A_i=\sum_{j\neq i} X_{ij} e_j^{\rho}$.
The map $f(e_i)=e_i^{\rho-1}-c_i(e_i^{\rho}+A_i)$ is strictly decreasing
on $\RR_{>0}$, so the best response is unique and obtained by a scalar
root-find; when $A_i=0$ it admits the closed form $e_i^*=1/c_i$.
 $\rho\in(0,1)$ sets the substitutability of neighbors'
efforts: values near $1$ make spillovers more additive, while smaller
values make one neighbor's protection less able to substitute for others'.
As $\rho\to1$, Eq.~\eqref{eq:util_ces} recovers log-linear case
in~\eqref{eq:util_log}. We use $\rho\in\{0.3,0.5,0.7\}$ to probe different
degrees of nonlinearity.%sets the substitutability of neighbors' efforts: values near $1$ make spillovers additive and highly substitutable, while smaller values make one neighbor's protection less able to substitute for others'. Equation~\eqref{eq:util_ces} is therefore a one-parameter generalization of the log-linear aggregate rather than an unrelated third utility, and $\rho\in\{0.3,0.5,0.7\}$ probes progressively more nonlinear aggregates within its defining range.

\smallskip \noindent \textbf{Research goal.}
We aim to learn an amortized solver that maps the parameters of an IDS game directly to its Nash equilibrium. Let $(X,c)$ denote a game instance, with the scalar $\rho$ additionally included for the log-CES utility family. We train a  predictor $f_\theta$ to approximate the equilibrium map by minimizing
$\E_{(X,c)\sim\mathcal D}\bigl[\norm{f_\theta(X,c)-e^*(X,c)}^2\bigr],$ where $e^*(X,c)$ is the Nash equilibrium of the game. We denote the predicted equilibrium by $\hat e=f_\theta(X,c)$.

% ==================================================================
\section{BRAID: Weight-Tied Iterative GNN}
\label{sec:model}
% ==================================================================

\begin{figure}[t]
\centering
\includegraphics[width=\textwidth]{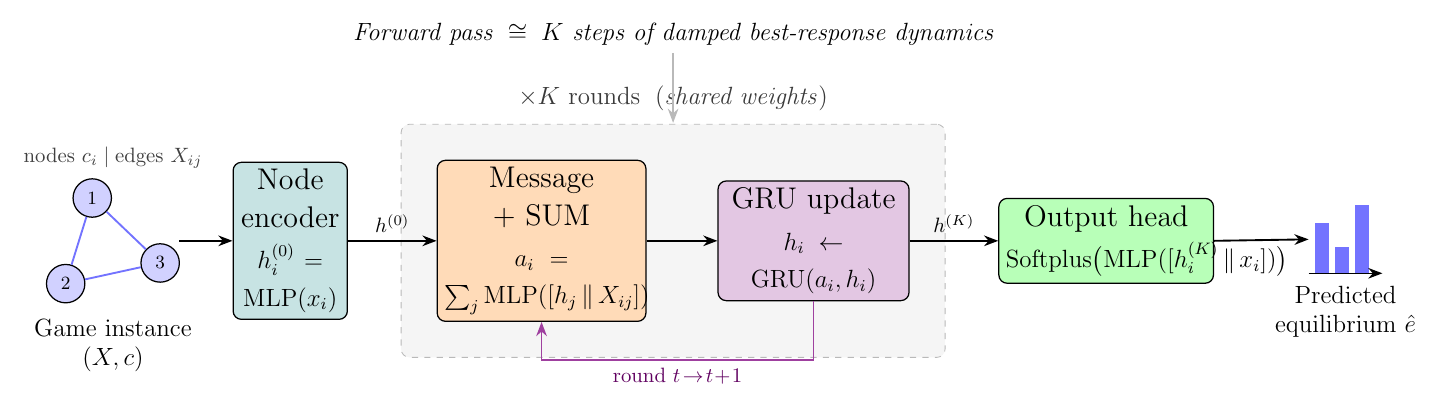}
\caption{Architecture of BRAID, our weight-tied iterative GNN. A game
  instance $(X,c)$ enters as a graph with node features $x_i=[c_i,X_{ii}]$
  and edge weights $X_{ij}$. A node encoder produces initial states
  $h^{(0)}$; the shared message-and-SUM block and GRU update are then
  applied for $K$ weight-tied rounds (the recurrence shown by the feedback
  arrow), so the forward pass mirrors $K$ steps of damped best-response
  dynamics. A Softplus output head maps the final states to the predicted
  equilibrium effort profile $\hat e$.}
\label{fig:architecture}
\end{figure}

\subsection{From Best-Response Dynamics to a Weight-Tied GNN}
\label{sec:motivation}

Across the utility specifications in Sec.~\ref{sec:utilities}, agent $i$'s
best response depends on the rest of the network through a weighted
neighbor aggregate:
\begin{equation}  \BR_i(e_{-i}) = g_i\!\bigl(S_i;\,c_i\bigr),
  \qquad
  S_i = \sum_{j\neq i} X_{ij}\,\phi(e_j),
  \label{eq:unified_br}
\end{equation}
where $\phi(e)=e$ for the log-linear and quadratic-cost utilities and
$\phi(e)=e^{\rho}$ for the log-CES utility.
Damped best-response dynamics defines the update operator  $T(e)  = (1-\alpha)e + \alpha\,g\!\bigl(S(e);c\bigr)$ and iterates
\begin{equation}
  e^{(t)} = T(e^{(t-1)}) = (1-\alpha)\,e^{(t-1)}
            + \alpha\,g\!\bigl(S(e^{(t-1)});\,c\bigr),
  \label{eq:brd_operator}
\end{equation}
until convergence. A Nash equilibrium is a fixed point, i.e., $e^*=T(e^*)$.

This structure suggests a natural graph neural network (GNN) architecture. Each agent's update is
local, depending on its own parameters and an additive, edge-weighted
summary of neighboring efforts, and the same update is applied at every
step of the dynamics. BRAID therefore uses a weight-tied iterative GNN:
SUM-aggregated message passing captures the weighted neighbor aggregate,
while shared parameters across rounds mirror the repeated application of
the same best-response operator. Concretely, a message-passing round computes
\begin{equation}
  m_i^{(t)} = \sum_{j\in\mathcal N(i)} X_{ij}\,h_j^{(t-1)},
  \label{eq:message}
\end{equation}
where $\mathcal N(i)=\{j\neq i:X_{ij}>0\}$ and $h_j^{(t-1)}$ is the
current hidden representation of agent $j$. The hidden state plays the
role of a learned effort representation. Replacing the map $g$ with a learned update cell, and applying the same
cell for $K$ rounds, yields a forward pass that mirrors $K$ damped
best-response steps.

\smallskip \noindent \textbf{Why Message Passing Fits IDS Games.}
\label{sec:whygnn}
Network game theory suggests that topology is
essential: an agent's equilibrium effort depends not only on its private
cost, but also on its structural position in the network
\cite{ballester2006s,bramoulle2014strategic,galeotti2010network}. In IDS
games, the graph is therefore not auxiliary information; it is the main
object through which strategic dependencies are expressed. A flat
predictor, such as a multilayer perceptron (MLP) applied to a fixed-length
encoding of $(X,c)$, does not encode
the structural properties of the equilibrium map: it is not permutation equivariant, so invariance to agent
relabeling must be learned rather than imposed; it has no built-in notion
of locality, so it cannot exploit graph sparsity; and it is tied to a fixed
number of agents.
Attention-based graph models, such as graph attention
networks \cite{velickovic2018gat},
restore equivariance but introduce an
additional learned weighting over neighbors, even though the
coupling weights $X_{ij}$ are already specified by the game. In contrast, edge-weighted SUM message passing uses the given edge
weights and directly matches the best-response aggregate $S_i$.
SUM aggregation also preserves additive multiplicity information that
mean or max aggregation can discard, and is known to be maximally
expressive among common message-passing aggregators \cite{xu2019gin}.
For these reasons, we use edge-weighted SUM message passing as the GNN
inductive bias most closely aligned with the fixed-point structure of IDS
best responses.

\smallskip \noindent \textbf{Why Prior Learned Solvers Do Not Directly Apply.}
\label{sec:whynot} The methods of Sec.~\ref{sec:related} do not transfer directly to
continuous-effort IDS games. Applying NES
\cite{marris2022turbocharging} would require discretizing each agent's
action space, producing a payoff tensor that grows exponentially with the
number of agents while obscuring the sparse network structure and
precluding sensitivity analysis in continuous parameters. N-FPNs
\cite{mckenzie2022operator} handle continuous games but through a generic
variational-inequality formulation, iterated to convergence at inference
time: they exploit neither graph sparsity nor additive neighbor coupling,
and provide no bounded-depth single-pass predictor. Making such a method
graph-aware and replacing convergence-time iteration with a fixed unroll
leads precisely to the weight-tied, SUM-aggregating design developed
here. RENES \cite{wang2024renes} targets finite-action normal-form games,
builds its graph over joint action profiles rather than the agent
topology, and still requires an inner solver at inference; the Exp3-IX
learner \cite{yu2024explore} learns online in a single environment and
targets the weaker CCE concept; and inverse-game methods such as MINE
\cite{trivedi2020mine} infer game parameters from observed equilibria
rather than predicting equilibria from known games. We address the
related non-smoothness of equilibrium maps from the forward direction, by
focusing training and evaluation on interior equilibria where the
best-response map is smooth (Sec.~\ref{sec:corner}).

In summary, existing learned solvers are not simultaneously forward,
graph-aware, single-pass, and designed for continuous-action network games.
BRAID is designed to satisfy these requirements by aligning the computation
of the neural solver with the fixed-point structure of IDS best-response
dynamics.

\subsection{Architecture Details}\label{sec:design}

BRAID has three core components (Fig.~\ref{fig:architecture}). First, we use \textbf{SUM aggregation}
because the best-response aggregate
$S_i$ in~\eqref{eq:unified_br} is additive and weighted by the edge intensities $X_{ij}$. Second, we use
\textbf{GRU gating} as a learned analogue of the damped update
$e\leftarrow(1-\alpha)e+\alpha\,\BR(e)$ in~\eqref{eq:brd_operator}, which
stabilizes iterative best-response dynamics. Third, we use
\textbf{weight tying} across all $K$ rounds to reflect the time-homogeneous
structure of the fixed-point operator. 
\begin{itemize}[leftmargin=*]
    \item 
  \textit{Node encoder.}
Each node has raw features $x_i=[c_i,\,X_{ii}]^\top$, consisting of its
cost and self-weight. These features are mapped to an initial hidden state
by a two-layer MLP with layer normalization:  $h_i^{(0)}=\text{MLP}_\text{enc}(x_i)\in\RR^d$.

\item  \textit{Message and aggregation.}
At each round $t\in\{1,\ldots,K\}$, agent $i$ receives messages from its
neighbors:
\begin{equation}
  \mu_{j\to i}^{(t)}
    = \text{MLP}_\text{msg}\!\bigl([h_j^{(t-1)}\,\|\,X_{ij}]\bigr),
  \qquad
  a_i^{(t)} = \sum_{j\in\mathcal N(i)} \mu_{j\to i}^{(t)},
\end{equation}
where $[\cdot\,\|\,\cdot]$ denotes concatenation and $X_{ij}$ is the scalar edge weight. The SUM aggregation mirrors the additive structure of the
best-response aggregate.

\item  \textit{GRU update with weight tying.} The hidden state of each node is then updated:
\begin{equation}
  h_i^{(t)} = \text{GRU}\!\bigl(a_i^{(t)},\;h_i^{(t-1)}\bigr),
  \label{eq:gru}
\end{equation}
using the same GRU  at every round $t$. This weight tying mirrors
the repeated application of the best-response operator in
damped best-response dynamics.

\item  \textit{Output head.}
After $K$ rounds, we get the effort
prediction:
\begin{equation}
  \hat e_i
    = \text{Softplus}\!\Bigl(\text{MLP}_\text{out}
        \!\bigl([h_i^{(K)}\,\|\,x_i]\bigr)\Bigr),
  \label{eq:output}
\end{equation}
where the raw node features $x_i=[c_i,X_{ii}]^\top$ are re-injected as a
skip connection so the output layer has direct access to the analytic
$1/c_i$ dependence, and Softplus enforces $\hat e_i>0$. \end{itemize}

\begin{algorithm}[t]
\caption{BRAID forward pass: single-pass equilibrium prediction. The $K$
  weight-tied rounds (lines 4--9) are isomorphic to $K$ damped
  best-response steps of the operator~\eqref{eq:brd_operator}, with the
  analytic best-response map replaced by learned message and update
  functions and the inner convergence loop replaced by a fixed, bounded
  depth.}
\label{alg:braid}
\begin{algorithmic}[1]
  \STATE \textbf{Input:} game $(X,c)$; node features $x_i=[c_i,X_{ii}]^\top$;
         neighbor sets $\mathcal N(i)$; rounds $K$
  \STATE \textbf{Shared parameters:} $\text{MLP}_\text{enc}$,
         $\text{MLP}_\text{msg}$, $\text{GRU}$, $\text{MLP}_\text{out}$. \textit{(tied across all rounds)}
  \STATE $h_i^{(0)} \leftarrow \text{MLP}_\text{enc}(x_i)$ for all $i$.
         \textit{(encode raw features)}
  \FOR{$t=1$ \TO $K$}
    \FOR{each agent $i$ in parallel}
      \STATE $a_i^{(t)} \leftarrow
        \sum_{j\in\mathcal N(i)}
        \text{MLP}_\text{msg}\!\bigl([h_j^{(t-1)}\,\|\,X_{ij}]\bigr)$. \textit{(edge-weighted aggregate $S_i$)}
      \STATE $h_i^{(t)} \leftarrow
        \text{GRU}\!\bigl(a_i^{(t)},\,h_i^{(t-1)}\bigr)$. \textit{(damped best-response update)}
    \ENDFOR
  \ENDFOR
  \STATE $\hat e_i \leftarrow
    \text{Softplus}\!\bigl(\text{MLP}_\text{out}([h_i^{(K)}\,\|\,x_i])\bigr)$
    for all $i$. \textit{(decode, $\hat e_i>0$)}
  \RETURN equilibrium prediction $\hat e=(\hat e_1,\ldots,\hat e_n)$.
       \textit{(one forward pass, no inner solve)}
\end{algorithmic}
\end{algorithm}

\noindent \textbf{Configuration.}
We scale the hidden width $d$ and the number of message-passing rounds
$K$ with the agent count: $(d,K)=(64,12)$ for the smallest games,
$(128,15)$ for $n=30$, and $(256,20)$ for $n\geq50$. For a fixed hidden
width, weight tying makes the parameter count independent of the unroll
depth $K$, so increasing the number of propagation rounds does not add
additional parameters. Table~\ref{tab:config} lists the full per-size
configuration.

\smallskip \noindent \textbf{Properties and complexity.}
BRAID is permutation equivariant with respect to agent relabeling: the
encoder, message function, GRU update, and output head are shared across
nodes, and the only neighborhood operation is SUM aggregation. For fixed
hidden dimension $d$ and unroll depth $K$, the parameter count is
independent of the number of agents. The computation in each round costs
 $O(\abs{\mathcal E}\,d + n\,d^2)$. Thus, the full forward
pass costs
$O\!\bigl(K(\abs{\mathcal E}\,d + n\,d^2)\bigr)$ which is linear in the number of edges for fixed $d$ and $K$. This bounded
forward pass replaces the instance-dependent and potentially unbounded
iteration count of best-response dynamics.

\smallskip \noindent \textbf{Relation to implicit and fixed-point networks.}
Deep equilibrium models \cite{bai2019deq} and Nash Fixed-Point Networks
\cite{mckenzie2022operator} define their output as the fixed point of a
single tied operator and compute that fixed point by iterating to convergence. This typically requires specialized training procedures such as implicit
differentiation or Jacobian-free backpropagation. BRAID instead unrolls the \emph{same}
tied operator for a fixed number of rounds, which can be viewed as a
finite-depth truncation of a fixed-point computation rather than a fully
converged implicit solve.
This has two practical consequences: the fixed unroll is an ordinary
computation graph, so standard backpropagation suffices; and the only
output constraint is positivity, enforced smoothly by the Softplus head,
which matters for recovering local sensitivities of the learned
equilibrium map (Sec.~\ref{sec:sensitivity}).

% ==================================================================
\section{Training}
\label{sec:training}
% ==================================================================

\begin{table}[t]
\centering
\caption{Graph and model configurations by network size. Graphs are
Erd\H{o}s--R\'enyi with edge probability $p$; $(d,K)$ denote the hidden
width and number of weight-tied rounds; lr is the base learning rate; bs
the batch size; and wd the weight decay.}
\label{tab:config}
\small
\setlength{\tabcolsep}{6pt}
\begin{tabular}{rcccccc}
\toprule
$n$ & $p$ & $d$ & $K$ & lr & bs & wd \\
\midrule
3   & 0.80 & 64  & 12 & $1\times10^{-3}$ & 32 & $1\times10^{-5}$ \\
30  & 0.15 & 128 & 15 & $8\times10^{-4}$ & 32 & $1\times10^{-5}$ \\
50  & 0.15 & 256 & 20 & $5\times10^{-4}$ & 32 & $1\times10^{-5}$ \\
100 & 0.15 & 256 & 20 & $3\times10^{-4}$ & 16 & $5\times10^{-6}$ \\
200 & 0.15 & 256 & 20 & $3\times10^{-4}$ & 16 & $5\times10^{-6}$ \\
\bottomrule
\end{tabular}
\end{table}

\noindent \textbf{Dataset Generation.}
\label{sec:dataset}
Each instance is a symmetrized Erd\H{o}s--R\'enyi graph on $n$ nodes with
self-benefit $X_{ii}=1$ and edge probability $p$ (Table~\ref{tab:config}).
For each game, we sample a coupling target $\tau\sim\mathcal U(0.3,1.2)$ and
set a base edge scale $b=\tau/[p(n-1)]$; edge weights are then drawn from a
two-component mixture, with $60\%$ from $\mathcal U(0.5b,b)$ (weak) and
$40\%$ from $\mathcal U(b,2b)$ (strong), so the expected neighbor aggregate
matches $\tau$ and the dataset does not collapse to a single coupling
regime. Costs are drawn as $c_i\sim\mathcal U(0.1,2.0)$ over a deliberately
wide range (Sec.~\ref{sec:corner}); for the log-CES experiments, a
separate dataset and model are trained for each $\rho\in\{0.3,0.5,0.7\}$.
Ground-truth equilibria are computed by damped BRD
(Algorithm~\ref{alg:brd}) with damping $\alpha=0.3$, tolerance $10^{-7}$,
initialization $e_i\sim\mathcal U(0,0.1)$, and up to $500$ iterations; each
returned profile is accepted only if its maximum best-response gap is below
$10^{-3}$. Instances for which BRD fails to converge or fails this
verification are re-drawn from the same distribution (up to five attempts
per game), rather than removed from a fixed pool, so the accepted set is
not biased toward structurally easier instances. We generate $10{,}000$
accepted instances per (utility, $n$) setting and use a $70/15/15$
train/validation/test split, holding out $1{,}500$ test games per setting.

%\begin{algorithm}[t]
%\begin{wrapfloat}{algorithm}{r}{0.53\textwidth}

\begin{algorithm}[t]
\caption{Damped Best-Response Dynamics (Ground Truth)}
\label{alg:brd}
\begin{algorithmic}[1]
  \STATE \textbf{Input:} $X\in\RR^{n\times n}$, $c\in\RR^n_{>0}$,
         utility-specific best response $\BR_i$
  \STATE \textbf{Parameters:} $T_\mathrm{max}$, tol, damping $\alpha$
  \STATE Initialize $e \sim \mathcal U(0,0.1)^n$
  \FOR{$t=1$ \TO $T_\mathrm{max}$}
    \STATE $e_\mathrm{prev}\leftarrow e$
    \FOR{$i=1$ \TO $n$}
      \STATE $\tilde e_i \leftarrow \BR_i(e_{-i})$
             \quad\textit{(Eq.~\eqref{eq:util_log},
             \eqref{eq:util_quad}, or \eqref{eq:util_ces})}
    \ENDFOR
    \STATE $e \leftarrow (1-\alpha)\,e + \alpha\,\tilde e$
    \IF{$\norm{e-e_\mathrm{prev}}_\infty < \text{tol}$}
      \RETURN $e$ \quad\textit{(converged)}
    \ENDIF
  \ENDFOR
  \RETURN $e$ \quad\textit{(did not converge)}
\end{algorithmic}
\end{algorithm}
%\end{wrapfloat}
%

\smallskip \noindent \textbf{Interior-Equilibrium Training.}
\label{sec:corner}
During development, we found that datasets with many boundary equilibria
($e_i^*=0$ for some agents) led to poor local behavior: the model could
achieve low prediction error while learning a nearly constant map in
regions where the best-response map is non-smooth in costs and edge
weights. Such regions are also unsuitable for sensitivity recovery, since
local derivatives can be ill-defined or dominated by active-set changes.
We therefore train and evaluate on an interior-equilibrium regime. The
\textit{cost range} and per-game \textit{coupling target} are chosen so that
equilibria remain in the interior regime; we monitor the fraction of
near-boundary efforts (those below $10^{-4}$) as a dataset statistic rather
than applying a hard post-hoc rejection threshold. Across the generated
datasets this fraction is negligible, keeping the target equilibrium map in
a smooth region of the best-response dynamics, which is also necessary for
stable finite-difference sensitivities.

\smallskip \noindent \textbf{Input Noise Regularization.}
To encourage local smoothness of the learned equilibrium map, we add
independent Gaussian noise with standard deviation $0.01$ to node and
edge features during training only. Perturbed edge weights are clamped to
remain non-negative. This acts as a data-space smoothness prior: if the
model produces very different predictions for nearly identical games, it
incurs larger expected MSE under the noisy inputs. No ground-truth
sensitivity labels are used.

\smallskip \noindent \textbf{Objective.}
For a single game instance, with prediction $\hat e=f_\theta(X,c)$, we
define the per-agent mean-squared error
$\ell(\theta;X,c)
    = \frac{1}{n}\sum_{i=1}^n \bigl(\hat e_i - e_i^*\bigr)^2 .$
The training objective is the expected loss over game instances,
\begin{equation}
  \mathcal L(\theta)
  =
  \E_{(X,c)\sim\mathcal D}
  \bigl[\ell(\theta;X,c)\bigr].
  \label{eq:loss}
\end{equation}
In practice, we minimize the empirical version of this objective over 
training set.
We deliberately do \emph{not} add a best-response-gap penalty to the
loss; instead we monitor the best-response gap (exploitability) as an
\emph{independent} held-out quality metric, so fixed-point
satisfaction is measured rather than fitted.
Optimization uses AdamW (base learning rate $10^{-3}$, reduced for larger
games as in Table~\ref{tab:config}; weight decay $10^{-5}$, reduced to
$5\times10^{-6}$ for $n>70$) with global-norm gradient clipping at
$1.0$, a ReduceLROnPlateau schedule (factor $0.5$) driven by
validation loss, and early stopping on validation MSE; training runs for
up to $300$ epochs.

% ==================================================================
\section{Sensitivity Analysis}
\label{sec:sensitivity}
% ==================================================================

Pointwise equilibrium accuracy does not by itself guarantee accurate
predictions of how equilibria change under parameter perturbations. A learned solver can achieve low prediction MSE at
sampled game instances while still producing incorrect local responses to
changes in costs or edge weights. Since such sensitivities are central to
auditing, stress testing, and intervention design, we evaluate whether
BRAID recovers the local sensitivity of the equilibrium map.

\paragraph{Finite-difference argument.}
Let $\gamma$ collect the scalar parameters of a game instance, including
costs and edge weights, and let $u_j$ denote the unit (basis) vector
corresponding to parameter $\gamma_j$. The true equilibrium map is
$\gamma\mapsto e^*(\gamma)$, and BRAID learns
$\gamma\mapsto f_\theta(\gamma)$. If the learned map is accurate not only
at a nominal point $\gamma_0$, but also throughout the perturbation
neighborhood $\gamma_0\pm\varepsilon u_j$, then substituting these
approximations into a symmetric finite difference gives
\begin{equation}
  \frac{f_\theta(\gamma_0+\varepsilon u_j)-f_\theta(\gamma_0-\varepsilon u_j)}
       {2\varepsilon}
  \;\approx\;
  \frac{e^*(\gamma_0+\varepsilon u_j)-e^*(\gamma_0-\varepsilon u_j)}
       {2\varepsilon},
\end{equation}
Thus, sensitivity recovery requires more than small error at
$\gamma_0$: the prediction error must remain small and smooth across the
finite-difference neighborhood, otherwise the difference of errors can be
amplified by the factor $1/(2\varepsilon)$. Interior-equilibrium training
keeps the target map away from non-smooth boundary responses, while
input-noise regularization discourages spurious local oscillations of the
learned map. Sec.~\ref{sec:experiments} evaluates this premise by
measuring prediction error at both nominal and perturbed games.

\paragraph{Protocol.}
For each test game $\gamma_0=(X_0,c_0)$ and each scalar parameter
$\gamma_j$ (cost $c_j$ or directed edge weight $X_{jk}$), with
corresponding basis vector $u_j$, we compute symmetric finite differences
for both the true and learned map:
\begin{align*}
  \frac{\partial e^*}{\partial\gamma_j}\bigg|_{\gamma_0}
    \approx \frac{e^*(\gamma_0+\varepsilon u_j)
                  - e^*(\gamma_0-\varepsilon u_j)}{2\varepsilon},~~~
  \frac{\partial f_\theta}{\partial\gamma_j}\bigg|_{\gamma_0}
    \approx \frac{f_\theta(\gamma_0+\varepsilon u_j)
                  - f_\theta(\gamma_0-\varepsilon u_j)}{2\varepsilon},
  \label{eq:fd_model}
\end{align*}
with $\varepsilon=10^{-2}$.
Ground-truth sensitivities require two extra BRD solves
per parameter; model sensitivities  require two extra
forward passes.
Because the ground-truth computation is
expensive, sensitivity metrics are averaged over a sampled subset of test
games for each setting, while equilibrium and perturbation errors are
computed on the full held-out set. Using symmetric differences on both
sides gives a consistent comparison.

\paragraph{Perturbed-equilibrium verification.}
When computing $e^*(\gamma_0\pm\varepsilon u_j)$, BRD is warm-started from
the unperturbed equilibrium $e^*(\gamma_0)$ rather than from a random
initialization, so the
solver stays in the same basin and the finite difference measures a true
local sensitivity rather than a discontinuous jump to a different
equilibrium.
We additionally re-verify each perturbed solution and discard any that
fails the equilibrium check.
Algorithm~\ref{alg:sensitivity} summarizes the full procedure.

\begin{algorithm}[t]
\caption{Sensitivity recovery: warm-started, verification-gated finite
  differences. Both the true and the learned Jacobian use symmetric
  differences; ground truth recomputes each perturbed equilibrium with a
  re-verified, warm-started best-response solve, so the difference measures
  a local sensitivity rather than a jump between equilibria.}
\label{alg:sensitivity}
\begin{algorithmic}[1]
  \STATE \textbf{Input:} test game $\gamma_0=(X_0,c_0)$ with equilibrium
         $e^*(\gamma_0)$; trained map $f_\theta$; step $\varepsilon=10^{-2}$;
         basis vectors $u_j$
  \FOR{each scalar parameter $\gamma_j$ (cost $c_j$ or edge weight $X_{jk}$)}
    \FOR{$s\in\{+\varepsilon,\,-\varepsilon\}$}
      \STATE $\gamma_s \leftarrow \gamma_0 + s\,u_j$
             \quad\textit{(perturb one coordinate)}
      \STATE $e^*(\gamma_s) \leftarrow \text{BRD}(\gamma_s)$ warm-started at
             $e^*(\gamma_0)$ \quad\textit{(stay in basin)}
      \IF{$e^*(\gamma_s)$ fails the equilibrium check}
        \STATE discard $\gamma_j$; \textbf{break}
      \ENDIF
      \STATE $\hat e(\gamma_s) \leftarrow f_\theta(\gamma_s)$
             \quad\textit{(one forward pass)}
    \ENDFOR
    \STATE $g_j^{*} \leftarrow
      \bigl[e^*(\gamma_0{+}\varepsilon u_j)-e^*(\gamma_0{-}\varepsilon u_j)\bigr]
      /(2\varepsilon)$ \quad\textit{(true sensitivity)}
    \STATE $\hat g_j \leftarrow
      \bigl[\hat e(\gamma_0{+}\varepsilon u_j)-\hat e(\gamma_0{-}\varepsilon u_j)\bigr]
      /(2\varepsilon)$ \quad\textit{(model sensitivity)}
    \STATE record relative error $\norm{\hat g_j-g_j^{*}}/\norm{g_j^{*}}$
  \ENDFOR
  \RETURN per-parameter sensitivity errors
\end{algorithmic}
\end{algorithm}
% ==================================================================
\section{Experiments}
\label{sec:experiments}
% ==================================================================

\smallskip \noindent \textbf{Evaluation Protocol.}
\label{sec:protocol}
All equilibrium and perturbation errors are averaged over the held-out
test set of $1{,}500$ games for each agent count $n$. Damped
best-response dynamics (BRD), iterated to convergence and re-verified, is
the ground-truth procedure: it provides the training targets and the
reference solutions against which BRAID's single-pass predictions are
evaluated. Alongside prediction error, we monitor the best-response gap
(exploitability) as an independent check of fixed-point satisfaction,
computed from quantities that are not directly optimized by the MSE loss.

 We report relative equilibrium error averaged over test games. Perturbation errors use the same metric, but
are evaluated at perturbed game parameters $\gamma_0\pm\Delta u_j$ with
$\Delta=10^{-2}$. Sensitivity errors compare the finite-difference
Jacobians of the learned and ground-truth equilibrium maps, as defined in
Sec.~\ref{sec:sensitivity}, and are averaged over sampled test games
because ground-truth sensitivities require additional BRD solves.

\begin{table}[t]
\centering
\caption{Equilibrium prediction, perturbation stability, and sensitivity
  recovery across utilities and agent counts ($\Delta=10^{-2}$).
  Equilibrium and perturbation errors are averaged over $1{,}500$ test
  games per $n$; sensitivities over a sample of test games per setting.
  The framework is identical across rows; only the best-response dynamics
  differ.}
\label{tab:size_gen}
\footnotesize
\setlength{\tabcolsep}{4.5pt}
\begin{tabular}{llccccc}
\toprule
\textbf{Utility} & $n$ & \textbf{Eq.\ err.}
  & \textbf{Pert.\ cost} & \textbf{Pert.\ edge}
  & \textbf{$\partial/\partial c_i$} & \textbf{$\partial/\partial X_{ij}$} \\
 & & (\%) & (\%) & (\%) & (\%) & (\%) \\
\midrule
\multirow{5}{*}{Log-linear}
  & $3$   & $0.33$ & $0.32$ & $0.34$ & $4.0$ & $11.6$ \\
  & $30$  & $0.35$ & $0.35$ & $0.36$ & $3.5$ & $11.0$ \\
  & $50$  & $0.38$ & $0.38$ & $0.39$ & $3.3$ & $11.4$ \\
  & $100$ & $0.29$ & $0.35$ & $0.36$ & $3.2$ & $13.2$ \\
  & $200$ & $0.27$ & $0.33$ & $0.34$ & $3.0$ & $13.6$ \\
\midrule
\multirow{5}{*}{Log-quadratic}
  & $3$   & $0.31$ & $0.30$ & $0.31$ & $5.0$ & $15.1$ \\
  & $30$  & $0.34$ & $0.34$ & $0.34$ & $3.1$ & $11.9$ \\
  & $50$  & $0.39$ & $0.39$ & $0.39$ & $2.7$ & $14.9$ \\
  & $100$ & $0.40$ & $0.40$ & $0.42$ & $2.4$ & $15.8$ \\
  & $200$ & $0.45$ & $0.46$ & $0.46$ & $2.2$ & $16.4$ \\
\midrule
\multirow{5}{*}{Log-CES ($\rho{=}0.3$)}
  & $3$   & $0.92$ & $1.11$ & $1.10$ & $6.92$ & $14.14$ \\
  & $30$  & $1.63$ & $1.67$ & $1.67$ & $5.70$ & $11.80$ \\
  & $50$  & $1.10$ & $1.05$ & $1.05$ & $4.87$ & $14.45$ \\
  & $100$ & $1.07$ & $1.11$ & $1.11$ & $4.60$ & $13.49$ \\
  & $200$ & $1.05$ & $1.09$ & $1.09$ & $4.26$ & $14.17$ \\
\midrule
\multirow{5}{*}{Log-CES ($\rho{=}0.5$)}
  & $3$   & $1.71$ & $2.15$ & $2.15$ & $5.69$ & $14.80$ \\
  & $30$  & $1.12$ & $1.15$ & $1.15$ & $5.10$ & $13.83$ \\
  & $50$  & $1.18$ & $1.25$ & $1.25$ & $6.28$ & $16.84$ \\
  & $100$ & $1.01$ & $1.04$ & $1.03$ & $6.41$ & $15.78$ \\
  & $200$ & $1.07$ & $1.43$ & $1.43$ & $6.88$ & $14.66$ \\
\midrule
\multirow{5}{*}{Log-CES ($\rho{=}0.7$)}
  & $3$   & $1.24$ & $1.22$ & $1.75$ & $4.83$ & $31.32$ \\
  & $30$  & $1.79$ & $2.25$ & $2.26$ & $7.25$ & $25.43$ \\
  & $50$  & $1.34$ & $2.35$ & $2.35$ & $3.97$ & $19.34$ \\
  & $100$ & $1.74$ & $2.74$ & $2.75$ & $6.41$ & $22.48$ \\
  & $200$ & $1.95$ & $2.84$ & $2.86$ & $5.32$ & $19.29$ \\
\bottomrule
\end{tabular}
\end{table}

\smallskip \noindent \textbf{Equilibrium Prediction.}
Table~\ref{tab:size_gen} reports the three utilities across agent counts
$n$ under an \emph{identical} architecture, training loop,
and sensitivity pipeline. Only the best-response module differs.
Three findings stand out.
First, accuracy is high and stable in $n$: relative equilibrium error
stays below $0.5\%$ for the log-linear and quadratic utilities and below
$2\%$ for the more challenging CES utility ($R^2>0.999$ throughout),
and remains in that band from $3$ to $200$ agents. Because a separate
model and configuration is trained at each size (Table~\ref{tab:config}),
this demonstrates consistent scalability across network sizes rather than
generalization of a single trained model to unseen sizes.
The CES utility is harder, as expected, because the per-neighbor
nonlinearity $e_j^{\rho}$ injects additional curvature into both the
equilibrium map and its derivatives.
Second, prediction error at perturbed parameters remains close to the
base equilibrium error. The columns ``Pert.\ cost'' and ``Pert.\ edge''
report errors at $\gamma_0\pm\Delta u_j$ for $\Delta=10^{-2}$. These
values are nearly identical to the nominal equilibrium errors for the
log-linear and quadratic-cost utilities, and remain of the same order for
the CES utilities. This verifies
Sec.~\ref{sec:sensitivity}: BRAID is accurate not only at the nominal
game instance, but also in the local neighborhoods used for
finite-difference sensitivity recovery.
Third, the sensitivity pattern (discussed next) is consistent across all
reported settings.
The CES utility was additionally evaluated at $n=5$ and $n=10$, with
equilibrium error $1.53\%$ and $1.88\%$ respectively, continuing the flat
trend.

\begin{figure}[t]
\centering
\begin{minipage}[t]{0.49\textwidth}
  \centering
  \includegraphics[width=\textwidth]{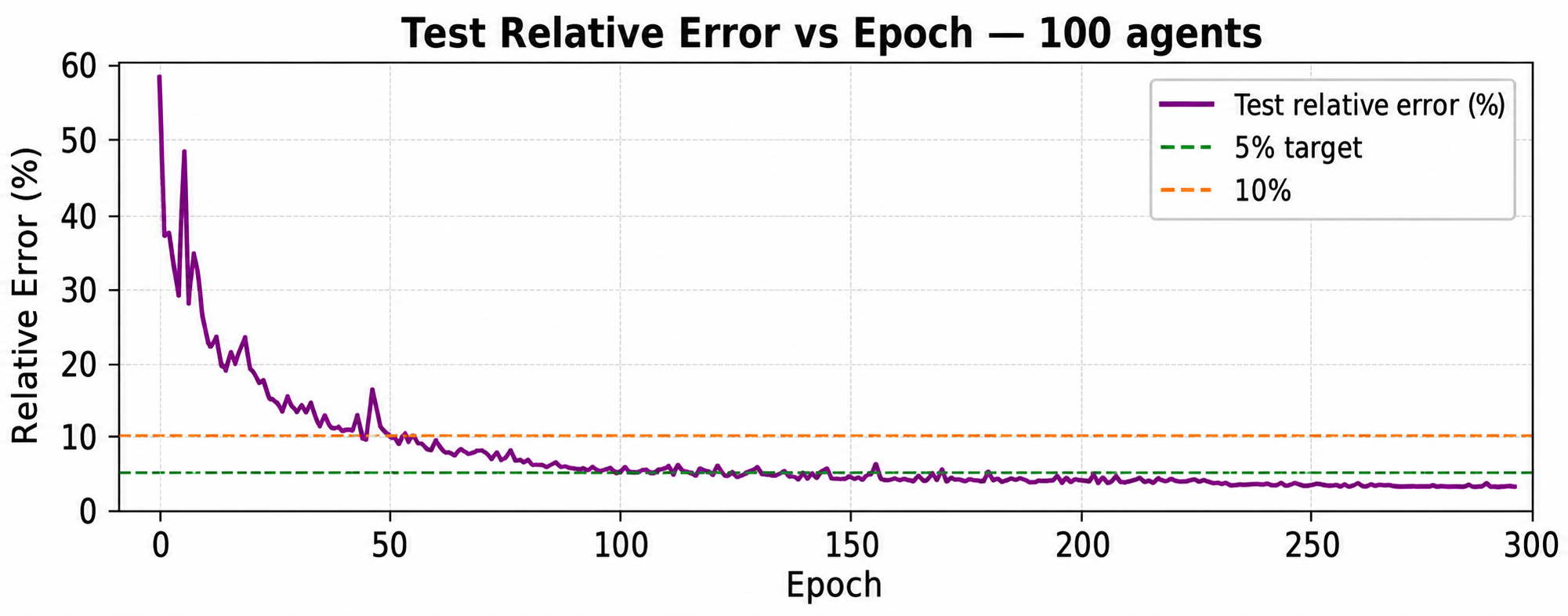}
\end{minipage}\hfill
\begin{minipage}[t]{0.49\textwidth}
   \raisebox{-2.3mm}{%
    \includegraphics[width=\textwidth]{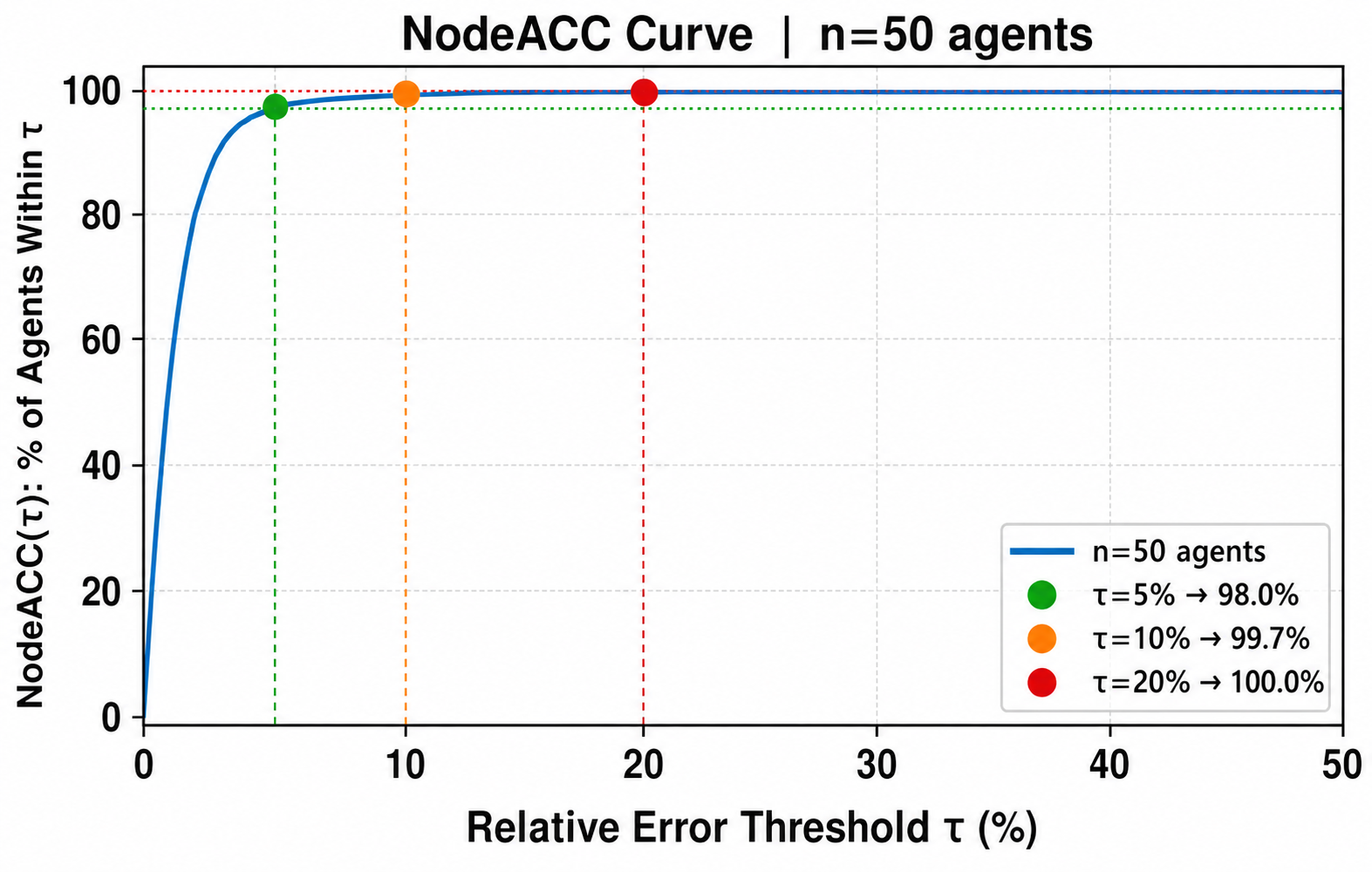}
}
\end{minipage}
\caption{\textit{Left:} test relative error versus training epoch for a
  representative $n=100$ model; it drops below the $5\%$ reference (green
  dashed) within about $40$ epochs and converges to $\sim\!0.5\%$ (orange
  dashed marks $10\%$), with no overfitting.
  \textit{Right:} per-agent NodeACC curve at $n=50$, the fraction of agents
  predicted within relative-error tolerance $\tau$ (a node-level Regression
  Error Characteristic curve \cite{bi2003rec}); at $\tau=5,10,20\%$ it
  reaches $98.0\%$, $99.7\%$, and $100\%$, confirming that accuracy is
  broadly uniform across agents rather than concentrated in an easy
  majority.}
\label{fig:results}
\end{figure}

\smallskip 
\noindent \textbf{Training dynamics.}
Training is stable across sizes.
Fig.~\ref{fig:results} (left) shows the test relative error during training
for a representative $n=100$ model: the error falls from roughly $58\%$ at
initialization, crosses the $5\%$ reference level within about $40$
epochs, and converges  to $\sim\!0.5\%$, with no sign of
divergence or overfitting.

\smallskip 
\noindent \textbf{Per-agent accuracy.}
The relative errors in Table~\ref{tab:size_gen} are averages, so they could
in principle hide a tail of poorly predicted agents.
To rule this out we report a per-agent \emph{NodeACC} curve, a node-level
Regression Error Characteristic curve \cite{bi2003rec}: for each tolerance
$\tau$ we plot the fraction of agents whose predicted effort lies within
relative error $\tau$ of the true equilibrium effort, which is the
empirical cumulative distribution of per-agent relative error.
Fig.~\ref{fig:results} (right) shows this curve for BRAID at $n=50$.
It rises almost vertically: $98\%$ of agents are predicted within $5\%$
relative error, $99.7\%$ within $10\%$, and all agents within $20\%$.
The accuracy is therefore broadly uniform across the network rather than
driven by a well-fit majority masking a poorly-fit tail.

\smallskip \noindent \textbf{Sensitivity Recovery.} The two sensitivity columns of Table~\ref{tab:size_gen} are the central
test motivated in Sec.~\ref{sec:sensitivity}.
Cost sensitivities $\partial e^*/\partial c_i$ are recovered with
$2$--$7\%$ relative error, while edge-weight sensitivities
$\partial e^*/\partial X_{ij}$ fall in the $11$--$17\%$ range for the
log-linear, quadratic, and $\rho\in\{0.3,0.5\}$ CES utilities, and rise to
$19$--$31\%$ for the most curved CES ($\rho=0.7$).
Edge sensitivities are thus consistently several times harder than cost
sensitivities across every utility and agent count, and the hardest
group throughout.
This asymmetry suggests it reflects how the two parameter types propagate
to the equilibrium rather than an artifact of one model: an edge weight
$X_{ij}$ influences the equilibrium only through an aggregate
($\sum_{j} X_{ij}e_j$ for U1/U2, $\sum_{j} X_{ij}e_j^{\rho}$ for U3), so
its sensitivity chain is more curved and many distinct edge perturbations
induce the same equilibrium change; the $\rho=0.7$ CES, whose aggregate is
the most sharply curved, is correspondingly the hardest case.
That the errors stay well below the level of a degenerate, near-constant
map is the evidence that the model has learned local structure, not merely
the conditional mean: when the training distribution places many
equilibria on the boundary (Sec.~\ref{sec:corner}), a model can obtain
modest equilibrium MSE while collapsing toward a nearly constant map whose
sensitivities are close to zero. Equilibrium MSE alone does not reveal
this degeneracy; sensitivity error does, which is why we treat it as a
first-class evaluation target.

\smallskip \noindent \textbf{Reliability of the BRD Ground Truth.}
\label{sec:reliability}
Because BRAID is trained and evaluated against BRD equilibria, we verify
that this ground truth is reliable and well-defined. First, non-convergent
instances are re-drawn during generation rather than dropped from a fixed
pool, and every retained profile is verified to a best-response gap below
$10^{-3}$, so the training targets are valid equilibria and the accepted
set is not biased toward easy games. Re-solving held-out games from a cold
random start, BRD converged within a $5{,}000$-iteration budget for most
instances, with the non-convergence rate growing with size
(Table~\ref{tab:runtime}, ``conv.''); the increasing but bounded iteration
count, not a failure to converge, is what motivates an amortized solver.
Second, to confirm BRAID matches a single well-defined equilibrium, we
re-solved each of $100$ held-out games from eleven initializations (the
stored equilibrium, six cold random starts, and four randomly perturbed
warm starts): at every network size all initializations converged to the
same profile, with a median maximum pairwise spread of
$\approx\!2\times10^{-6}$ and a maximum below $6\times10^{-6}$ (i.e., at
the solver tolerance), indicating an empirically unique equilibrium in the
tested regime. Finally, warm-started and cold-started perturbed solves
agreed in $100\%$ of $120$ perturbation tests per size, so the warm start
in Algorithm~\ref{alg:sensitivity} follows the same equilibrium branch
rather than manufacturing the measured sensitivity.

\begin{table}[t]
\centering
\caption{Runtime comparison on a single NVIDIA RTX A6000 (CUDA 12.1,
PyTorch 2.5.1, PyG 2.6.1) for the log-CES utility ($\rho=0.7$), the
hardest utility for BRD. BRD and BRAID use PyTorch implementations, with
BRD vectorized across agents. ``BRD iters'' reports the median (maximum)
number of damped best-response iterations to convergence among converged
instances; ``conv.'' is the fraction reaching the $10^{-7}$ tolerance
within a $5{,}000$-iteration budget. Latencies are medians at batch size
1. Speedups are BRD$_\text{cpu}$/BRAID$_\text{cpu}$ (same device) and
BRD$_\text{cpu}$/BRAID$_\text{gpu}$ (GPU deployment).}
\label{tab:runtime}
\small
\setlength{\tabcolsep}{5pt}
\begin{tabular}{rcccccc}
\toprule
$n$ & BRD iters & conv. & BRD$_\text{cpu}$ & BRAID$_\text{cpu}$ & BRAID$_\text{gpu}$
    & speedup \\
    & med (max) & (\%) & (ms) & (ms) & (ms) & (CPU / GPU) \\
\midrule
3   & 58 (97) & 100 & 112.1 & 3.34  & 2.59 & $34\times$ / $43\times$ \\
30  & 64 (83) & 95  & 128.7 & 6.41  & 3.48 & $20\times$ / $37\times$ \\
50  & 59 (73) & 90  & 119.7 & 13.36 & 4.40 & $9\times$  / $27\times$ \\
100 & 55 (65) & 78  & 115.1 & 20.95 & 4.63 & $5.5\times$ / $25\times$ \\
\bottomrule
\end{tabular}
\end{table}

\smallskip \noindent \textbf{Computational Performance.}
All timing uses a single NVIDIA RTX A6000 (CUDA 12.1, PyTorch 2.5.1,
PyG 2.6.1). To isolate the algorithm from implementation overhead, BRD is
vectorized across agents rather than looped, uses damping $0.3$ and
tolerance $10^{-7}$, and evaluates the CES best response by $60$ bisection
steps per agent per round. Table~\ref{tab:runtime} reports results for
log-CES ($\rho=0.7$), which is the worst case for BRD among our utilities:
its best response has no closed form and must be solved by scalar
root-finding at every agent and iteration, whereas the log-linear and
quadratic-cost best responses are closed-form and converge at least as
reliably. BRD requires a median of $55$--$64$ iterations (maximum $97$) at
an instance-dependent, unbounded cost: within a $5{,}000$-iteration budget
the non-convergence rate rises from $0\%$ at $n=3$ to $22\%$ at $n=100$.
BRAID replaces this loop with a fixed $K$-round forward pass costing
$2.6$--$4.6$\,ms on the GPU, a $25$--$43\times$ per-instance reduction
under GPU deployment and $5.5$--$34\times$ with both methods on the CPU;
the closed-form utilities show the same qualitative pattern. The advantage
compounds under batching (BRAID reaches $\sim\!10^{5}$ games/s at $n=3$)
and in sensitivity analysis, where each ground-truth finite difference
needs two additional BRD solves per parameter whereas BRAID needs only two
additional forward passes.

% ==================================================================
\section{Conclusion}
\label{sec:discussion}\label{sec:conclusion}
% ==================================================================
We presented BRAID, a weight-tied iterative graph neural network for Nash
equilibrium prediction in interdependent security games. Derived from the
best-response fixed point, BRAID uses weighted SUM aggregation and a shared
GRU update to mirror damped best-response dynamics. The same architecture
and training recipe apply across log-linear, quadratic-cost, and log-CES
utilities. BRAID achieves low equilibrium error across agent counts, while
interior-equilibrium training and input-noise regularization improve the
local behavior of the learned equilibrium map without sensitivity labels.
Using a verification-gated finite-difference protocol, we further show that
BRAID recovers meaningful cost and edge-weight sensitivities and matches a
BRD ground truth that is convergent and empirically unique in the tested
regime.

Several directions remain open. Edge-weight sensitivities remain the hardest
case, especially for log-CES with $\rho=0.7$, motivating richer edge
representations and edge-focused regularization. The framework could also
extend to behavioral probability-weighting utilities
\cite{hota2018interdependent} and Stackelberg or repeated-game settings
\cite{huang2023stackelberg}. More broadly, BRAID illustrates how
learning-based solvers can complement iterative methods with fast amortized
equilibrium predictions and informative local sensitivities.

%Several directions remain open. Edge-weight sensitivities are the hardest case, especially for the more curved log-CES utility with $\rho=0.7$, suggesting richer edge representations, edge-focused regularization, or noise schedules as useful next steps. The same framework could also be extended to behavioral probability-weighting utilities \cite{hota2018interdependent} and to Stackelberg or repeated-game settings, where a fast differentiable surrogate would be useful for anticipating followers' equilibrium responses \cite{huang2023stackelberg}. More broadly, our results suggest that learning-based solvers can complement iterative methods by providing fast amortized equilibrium predictions together with informative local sensitivities.

\medskip
\noindent \textbf{Acknowledgments.}
This work was funded in part by the National Science Foundation under award numbers IIS-2202699, IIS-2416895, IIS-2301599, CMMI-2301601, DMS-2529302.

\medskip
\noindent \textbf{Disclosure of Interests.}
The authors have no competing interests to declare that are relevant to the content of this article.

% ==================================================================
\bibliographystyle{splncs04}
\bibliography{references}
% ==================================================================

\end{document}